\documentclass{aa}
\begin{document} 
\input psfig.tex 
\title{Galaxy luminosity evolution: how much is due to a model choice?} 
\author{S. Andreon} 
\institute{INAF--Osservatorio Astronomico di Brera, Milano, Italy}
\titlerunning{Galaxy luminosity evolution} 
\date{A\&A, in press} 
\abstract{
The cluster and field luminosity functions (LFs) determined on large homogeneous
samples ($N>2200$ galaxies each) are almost indistinguishable,
down to $M^* +4$ in the $r$ and $i$ filters, 
hence suggesting that the effect of the cluster environment
on the galaxy properties does not affect the galaxy luminosity function in red
bands. The similarity of the red band LFs in different environments suggests that
the galaxy mass function is preserved during the galaxy infall in the cluster. By
analyzing a large sample of galaxies in clusters, ideal from many points of view
(multicolor data, large size, many clusters, metric magnitudes) we found that
luminosity evolution is required by the data but the latter do not unambiguously
derive its flavour if a differential luminosity evolution between bright and
faint galaxies is allowed. We show that the LF parameters (slope, characteristic
magnitude and their evolution) and errors depend on assumptions in a way seldom
recognized in literature. We also point out logical inconsistencies between
hypothesis assumed in deriving literature LF and presented results, 
suggesting caution in interpreting similar published results.  
}   \maketitle 

\begin{figure*}
\centerline{%
\psfig{figure=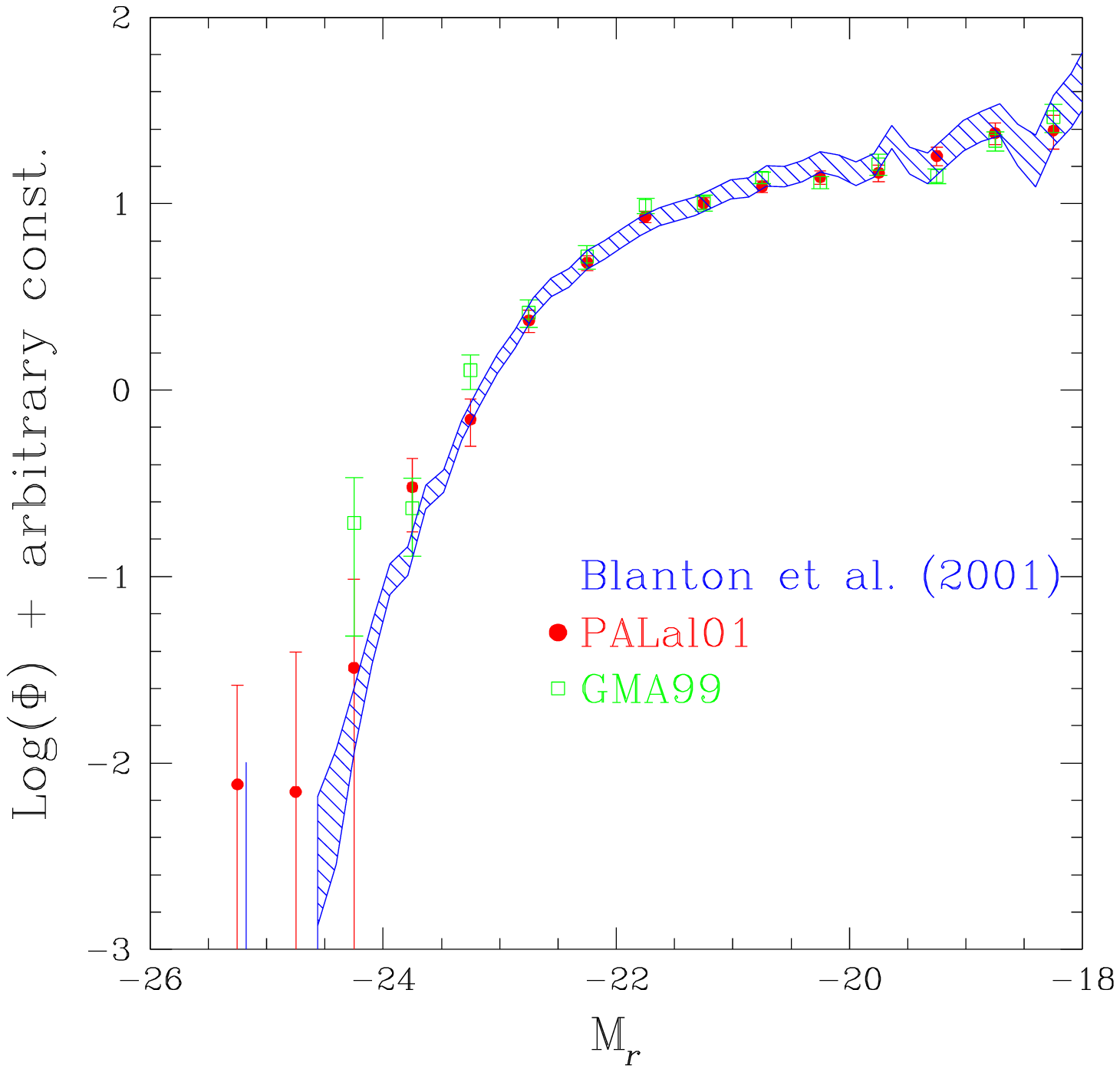,width=8.5truecm}
\psfig{figure=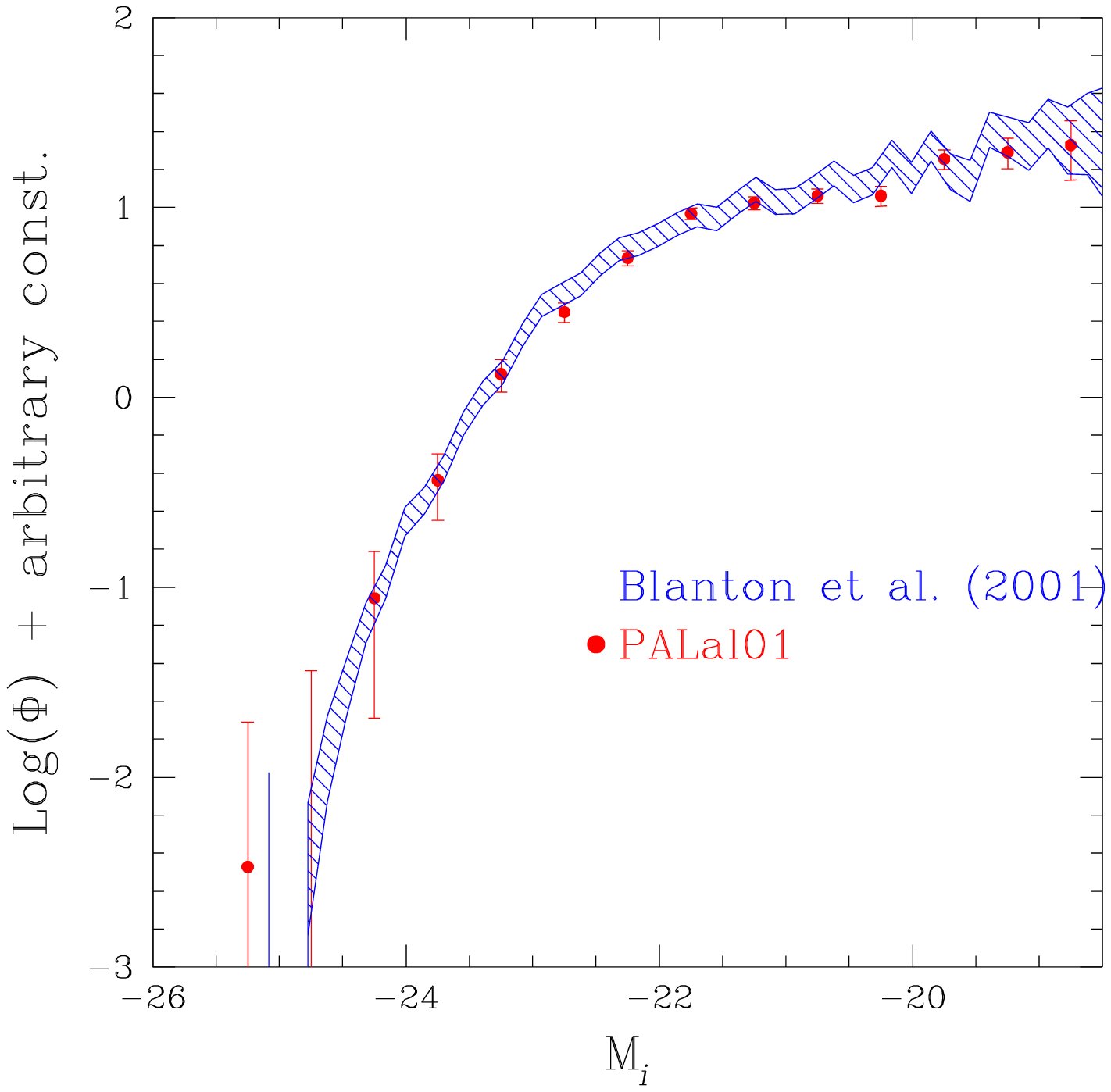,width=8.5truecm}}
\caption[h]{Luminosity function in the Gunn $r$ (left panel) and $i$ 
(right panel) for
the cluster and field environments. The shaded area is the field LF
derived by Blanton et al. (2001), green open and close red points mark
GMA99 and PALal01 LFs, respectively. GMA99 do not compute
any $i$ band LF using pseudo--total magnitudes.}
\end{figure*}

\section{Introduction}

The study of the cluster luminosity function (LF) has at least two immediate
objectives: to look for environmental related effects, as remarked
by possible differences between the cluster and field LF, and to look
for galaxy evolution, by comparing the LF at different redshifts.
Both the objectives are considered in this paper.

At the present date, cluster and field LFs of large samples of galaxies in
the nearby universe ($z<0.25$) have been computed (Garilli, Maccagni \& Andreon
1999, hereafter GMA99; Blanton et al. 2001;  Paolillo, Andreon, Longo et al.
2001,  hereafter PALal01; Norberg et al. 2002b; de Propris et al. 2003). A
convergence on the field LF, absent just few years ago (GMA99, see their figure
10) seems to be reached (Norberg et al.  2002b), in spite of an initial 
disagreement (Blanton et al. 2001) between Sloan Digital Sky Survey (SDSS) 
and 2 degree Field Galaxy Redshift Survey (2dFGRS) LFs. A convergence on the
cluster LF, when the considered sample is large enough to dump out possible
vagaries of individual cluster LF, seems also reached, given the good agreement
between the GMA99 LF, based on 65 clusters, the PAal01 LF, computed for 39
clusters, and older LFs (see GMA99 and PALal01 for details). There are,
however, recent cluster LFs that systematically differ from all LFs published
thus far (e.g. Goto et al. 2002).

Given the reached convergence on the field and cluster LFs, 
it is therefore time to compare them.

LFs are usually characterized by a Schechter (1976) function:

$$ \phi(M)=\phi^*10^{0.4(\alpha+1)(M^*_0-M)}e^{-10^{0.4(M^*-M)}} $$

where $M^*$ and $\alpha$ are the characteristic magnitudes and the
slope of the function, respectively. $\phi^*$ is a normalization parameter
not relevant in this work.

For the {\it field} environment, the evolution of the LF, i.e the $z$ dependence of
the best fit parameters of the LF, has been explored. Lin et al. (1999) presented
some evidence for an evolution on $M^*$ in the redshift range $0.12<z<0.55$, of the
order of 0.4 (0.3) mag in the rest--frame $B$ ($R$) over the explored redshift range.
Very recently, Blanton et al. (2003) analysis of a larger SDSS data set found a
similar variation for $M^*$ in the redshift range $0.02<z<0.22$ in $g',r',i'$.
Blanton et al. (2003) and Lin et al. (1999), both claim evolution {\it assuming} a
not evolving LF shape.

For the clusters, evolution of the LF is, instead, much less explored. GMA99
found only marginal, if any, evidence for evolution in the redshift range
$0.05<z<0.25$ by binning cluster LFs in two $z$ ranges. de Propris et al.
(1999) found evolution on $M^*$ in the $K$  band in the $0.2<z<0.9$ redshift
range for $M<M^*+1$ galaxies and assuming that $\alpha$ does not evolve
with $z$.

In this paper we have a twofold objective: first to compare the cluster and field
LFs (\S 2), second to measure the evolution of the cluster LF {\it
using the same formalism used for measuring the evolution of the field LF}.
 The data and the
analysis is summarized in \S 3, whereas results are presented in Section 4.

We adopt $H_0=50$ km s$^{-1}$ Mpc$^{-1}$ and $\Omega=1$. 
However, the $H_0$ value reduce off in the comparisons, and is, therefore,
irrelevant. Furthermore, the considered samples are at low redshift.

\section{Cluster and field LF}

The comparison of the cluster and field LF have been attempted several
times in literature (for example Bromley et al. 1998a,b; Christlein 2000; 
Balogh et al. 2001; Valotto et al. 1997; 
Trentham \& Hodgkin 2002), with opposite claims about
the LF environmental dependence.

Figure 1 shows three luminosity functions in the Gunn
$r$ and $i$ bands of the largest and most homogeneous samples
measured thus far. Green open points mark the
GMA99 cluster $r$ band LF. It has been obtained in the Gunn $r$
filter for about 2200 galaxies in 65 clusters, ranging in redshift
from 0.05 to 0.25. Red closed points mark 
the PALal01 $r$ and $i$ band LFs. The LFs
have been obtained by using $F$ and $N$ photographic plates
calibrated in the $r$ and $i$ Gunn photometric system. The sample
is about 1.5 time larger than the GMA99 one and is drawn from 39
clusters in the redshift range $0.06<z<0.28$. 
The two cluster LFs shown in
Figure 1 are derived using different materials (CCD vs
photographic plates), different ways to subtract 
interloper galaxies (using colors {\it vs} using control fields), 
different definitions of ``total" magnitude, different
portions of the clusters (GMA99 explored only the center, whereas
PALal01 considered the whole cluster), and finally, the absolute
limiting  magnitude is approximately independent on $z$ for
GMA99, whereas it is systematically 
brighter for more distant clusters in PALal01. Although
some clusters are in common between the two works, only a very minor
fraction of 
galaxies are in common, because of the different depths
and field of views of the two works. Therefore, the two
LFs are really independent each other, and should not
share common biases introduced by the data analysis (that
is completely different in the two works).

The dashed area in Figure 1 show the $r'$ and
$i'$ field LF, as measured by the SDSS (Blanton et al. 2001). The
sample is larger than both previous samples by a factor of few (at most), and
it extends over a similar redshift range: $0.02<z<0.17-0.20$. SDSS $r'$ and $i'$ 
filters are
quite similar to the Gunn $r$ and $i$ filters, at least for
galaxies at low redshift, because they map the emission coming from
almost identical part of the galaxy spectrum. The major difference
between the two systems is the star that defines the objects
of zero magnitude (and color). In fact, for $r$ and $i$ filters,
the conversion from SDSS to Gunn
is almost independent on the object spectra (see Table 3
in Fukugita et al. 1995). We therefore correct the SDSS
$r'$ and $i'$ SDSS magnitudes for the change of the zero mag standard star
by using the values listed in Fukugita et al. (1995).
Galaxies having a surface
brightness profile following a de Vaucouleurs (1948) law have a
Petrosian flux  that underestimates by 0.1-0.2 mag
the total mag (Blanton et al. 2001). Since most of the galaxies in cluster are
early--type galaxies, and hence have a de Vaucouleurs (1948) radial
profile, we made SDSS Petrosian magnitudes brighter by 0.15 mag, in order
to make them comparable to ``total" magnitudes derived in GMA99 
(SExtractor (Bertin \& Arnouts (1996) isophotal corrected
magnitudes) and in PALal01 (Focas (Jarvis \& Tyson 1981) ``total" magnitudes).
Finally, SDSS absolute magnitudes are corrected to our value of $H_0$. 
The vertical normalization of the LFs is arbitrary. 

The three LFs are almost indistinguishable, hence suggesting that
the effect of the cluster environment on the galaxy properties does
not affect the galaxy luminosity function in red filters. 
The comparison performed
by de Propris et al. (2003) between cluster and field $B_j$ LF
shows instead differences in
the cluster and field {\it blue} LF. The similarity of the luminosity
function in different environments in the red bands suggests that the galaxy
mass function is preserved during the galaxy infall in the cluster, whereas
the blue luminosity, more sensitive to star bursts, is altered.

The found similarity of the LF in different environments seems at 
variance with previous (opposite) findings. 

-- Many LF determinations (e.g. 
Small, Sargent \& Hamilton 1997; Bromley et al. 1998b; Christlein 2000; 
Balogh et al. 2001) use Sandage, Tammann \& Yahil (1979, STY)
and Efstathiou, Ellis \& Peterson (1988, EEP) LF estimators, in
which the spatial dependent part of the LF
factors out under the assumption of an LF universality\footnote{See, for details, EEP
and Binggeli, Sandage \& Tammann 1988}. These works
claim to have measured different LFs in different environments, in
spite of having assumed that the LF does not depend on 
environment in the LF computation.  Hypothesis and results are, therefore, 
in direct contradiction.
Furthermore, according to Efstathiou, Ellis \& Peterson 
(1988), STY and EEP estimators
underestimate errors if the LF universality does not hold.
Because of the logical 
inconsistency between hypothesis and conclusions, 
we consider the results of these papers as suggestive
at most, and with underestimated errors. 

We also note that the skewed (Blanton et al. 2001) LF of the Las 
Campanas Redshift Survey undermines the claims of environmental effects 
based on that survey, such as Christlein (2000), Bromley et al. (1998b) and
Balogh et al. (2001) results. Furthermore, Balogh et al. (2001) 
uses the 2MASS photometry (Jarrett et al. 2000), criticized by
Andreon (2002a) because 2MASS misses dim galaxies and the outer halos
of normal galaxies, hence giving skewed LFs, as confirmed by Blanton
et al. (2003) and Bell et al. (2003).

-- Cluster and field are found sometime to have different LFs. Some 
examples are shown in Valotto et al. (1997) and Zabludoff \& Mulchaey (2000).
Almost none of the papers comparing cluster and field LFs 
uses homogeneous and large samples as the ones used in this
paper (see de Propris et al. 2003 as an exception), 
and none in the red bands considered in the present paper. 
Furthermore, only recently (Norberg et al. 2002b) there has been a convergence
on the field LF.  Therefore, the results of previous cluster {\it vs} field LF
comparisons largely rely on which field LF has been considered.
Just as an example in red bands, Zabludoff \& Mulchaey (2000) do find
differences between the LF of poor groups and the field when they adopt
for the field the Lin et al. (1996) LF. The latter LF differ systematically from the
SDSS LF for
the reasons described in Blanton et al. (2001). The Zabludoff \& Mulchaey (2000)
result does not tell us about environmental differences, but
about the slow convergence of the field LF to the present day determination.

\medskip

The similarity of the global {\it red} LFs in different environments is not
in contradiction with the suggested universality of the 
LFs of morphological types in the {\it blue} band (Sandage, Binggeli, \& 
Tammann 1985; Jerjen \& Tammann 1997; Andreon 1998). First, the LF of
the types are measured in filters different from the ones studied in this 
paper. Second, the simplest interpretation of the suggested
universality of the LFs of
the morphological types is that galaxies change neither type nor 
luminosity. However, the same data can be interpreted as galaxies 
change type and luminosity, preserving at the same time 
the blue LFs of morphological types and the global mass function. Such an
interpretations allows both the blue LF of the morphological types
and the global red LF to be universal. It naturally allows the blue
LF of cluster and field to be different, as observed by de Propris et al. (2003).
Independently of the correct interpretation of the observations,
the measure of the global mass function, here approximated by
the red LF, is important in order to know whether galaxies change their mass
during the infall in the cluster. Our present measure
suggests that mass is preserved, at least in a statistical sense.

We caution the reader not to compare the SDSS and PALal01 $g$ band LFs,
since the latter LF, although being internally consistent, is in a photometric
system that can be linked to the standard SDSS $g'$ system in a quite
complicate way.  We are now re-deriving the blue cluster LF, for a twice
larger sample, in its ``natural" system (i.e. in the $B_J$) (Paolillo et
al. 2004).

A technical detail should now be mentioned. All three LFs shown in Figure 1
are derived assuming that luminosity evolution in the studied redshift range is
negligible. Since all three LFs considered almost the same redshift range, the
comparison makes sense even in the presence of luminosity evolution in the
studied redshift range. In that case, the derived LFs are computed at the
median $z$, which is almost the same for the three LFs ($0.15, 0.12, 0.10$
for GMA99, PALal01 and Blanton et al. 2001, respectively).
Although this point is quite technical, this is an important one: the measured
LF depends on the model used in its derivations, as shown in Section 4. 
Our comparison uses LFs derived adopting the same model.

\section{Cluster LF evolution: data and analysis}

In order to measure the LF evolution
we use the sample presented in Garilli et al. (1996)
and whose LFs have been previously presented in GMA99. 
This sample is ideal for
such evolutionary study for several reasons. First,  because all
clusters are sampled at $M^*+3$ to $M^*+4$  
independently on redshift (see GMA99 figure 5).
The sample depth allows to measure the evolution of both $M^*$ and $\alpha$ at
all studied redshifts, {\it for the first time}. Second, the fact that both
$M^*$ and $\alpha$ can be measured at all studied redshifts should clarify the usual
degeneracy between the two parameters 
and makes the results less sensitive to mis--interpretations.
Third, GMA99 compute aperture magnitudes
fixed in the galaxy frame (they use a 20 kpc aperture), magnitude
that we adopted here. Such a metric
aperture does not suffer the problem of isophotal magnitudes, or
even pseudo--total one according to Dalcanton (1998), that 
bias the LF with $z$ in such a way to mimic evolution. The
choice of a metric aperture is therefore essential. However,
LFs derived with such an aperture are tilted with respect to
the one computed using pseudo--total magnitudes, and the former
cannot be compared to the latter. For this reason 
in Figure 1 we plot the GMA99 LF derived using pseudo--total magnitudes
and not the one computed using aperture magnitudes.
Fourth, three colors (Gunn $g$, $r$ and $i$) photometry is
available, allowing to study the wavelength dependence of the
possible evolution.

The only difference with GMA99 sample is that we discarded
one cluster (MS0013+1558), because the studied part of it
includes just one single galaxy. We remind that absolute
magnitudes are k--corrected assuming an elliptical spectrum,
with k--corrections taken from Frei \& Gunn (1994). 

We compute the LF using standard maximum--likelihood
methods (STY, EEP) used for the field LF determination.

Given a sample of N galaxies in M clusters at redshift
$z_1,z_2, ..., z_M$, we compute the likelihood $L$ by
the formula:

$$ \ln L = \sum_{clusters} \sum_{galaxies} p_i $$

where $p_i$ is the individual conditional probability

$$ p_i=p(M_i | z_i) = \frac{\phi(M_i)}{\int \phi(M)dM} $$

where the integral is evaluated over the range
$[-\infty, mag_{lim}]$, where $mag_{lim}$ is the limiting
magnitude, and where $\phi$ is the Schechter (1976) function
modified for allowing $\alpha$ and $M^*$ to vary with $z$:

$$ \phi(M)=\phi^*10^{0.4(\alpha_z+1)(M^*_z-M)}e^{-10^{0.4(M^*_z-M)}} $$

$$ \alpha_z=\alpha_0-Pz $$

$$ M^*_z=M^*_0-Qz$$

We hence take $M^*$ and $\alpha$ to vary linearly with $z$ with a
rate given by $Q$, and $P$, respectively. Positive values
for $P$ and $Q$ mean that in the past galaxies were brighter 
and the LF was steeper. 
Our $Q$ has the same meaning of $Q$ in  Lin et al.
(1999) and Blanton et al. (2003). Instead,
their $P$ parameter has a different
meaning from our $P$: we model the $\alpha$
evolution, whereas they model the $\phi^*$ evolution.

Best fit parameters are derived by
maximizing the likelihood.
Error ellipses in the ($M^*,\alpha$) or ($P,Q$) planes
are computed by finding the contour corresponding
to 

$$ ln L = ln L - \frac{1}{2} \Delta\chi^2 $$

where $\Delta \chi^2$ is the change in $\chi^2$ appropriate for the desired
confidence level and a $\chi^2$ distribution (with two degree of
freedom/interesting parameters for confidence contours, and one
degree of freedom for single parameters).

The computation of the LF parameters by using the
maximum likelihood method has a great advantage:
the $\phi^*_i$ parameters (one per cluster) 
cancel out in computing $p_i$
(see the $p_i$ equation), and the dimension of the 
overall space to be explored decreases from 68 to 4.

\begin{figure}
\psfig{figure=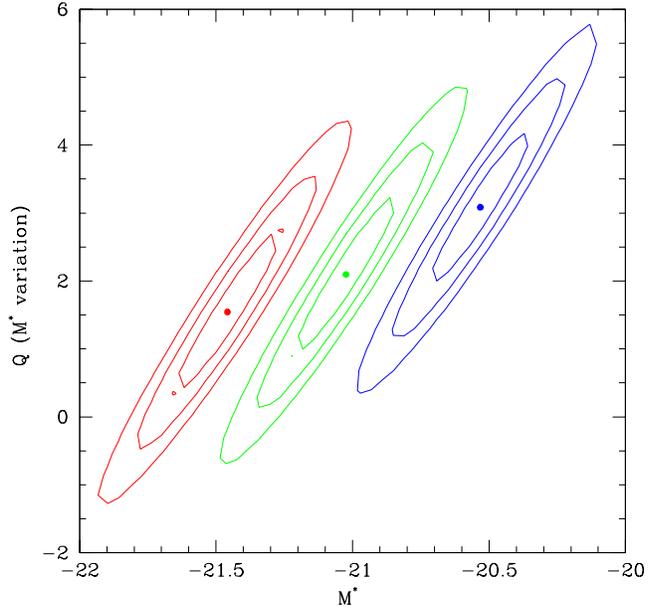,width=8.5truecm}
\caption[h]{Confidence contours at the 
68.3, 95.4 and 99.73 \% confidence levels 
for two degree of freedom for $M^*$ and $Q=\partial M^* / \partial z$ when 
$\alpha$ is not allowed to evolve. The no--evolution locus is given
by $Q=0$. For a passive evolving population $Q\sim1.1-1.2$.
 $g$, $r$ and $i$ contours are in blue, green and red
(also right to left). }
\end{figure}

\section{Genuine evolution or built-in the LF model?}

In literature, LFs have been computed under one of these
three hypothesis:

--  no evolution at all ($P=Q=0$). Only a few selected studies
do not make such an assumption. 

--  no evolution on the LF slope ($P=0$) and an assumed  
evolution on $M^*$. This hypothesis is used 
by the 2dFGRS LF (Norberg et al. 2002b).

--  no evolution on the LF slope ($P=0$). $M^*$ evolution ($Q$) is
solved during the LF determination (CNOC LF, Lin et al. 1999; SDSS LF, 
Blanton et al. 2003)

We determine the cluster LF under all these hypothesis (and even more).
Table 1 presents the derived best fit values and their errors.
We will show that results,
for the very same data sample, depend on the model used
to derive the LF, i.e. on assumptions about $P$ and $Q$. 
In particular, best fit parameters and their error,
luminosity evolution, and its statistical significance, all
depend on the adopted model, suggesting caution in
interpreting similar published results.

\subsection{The usual case, i.e. assuming $P=Q=0$}

By fixing $P=Q=0$,
i.e. assuming no evolution at all, we found the same best fit
($M^*, \alpha$) parameters found in GMA99, comfortably
showing that the present algorithm
converges to the same value previously found with an algorithm
using the $\chi^2$.
Presently and previously computed confidence contours also almost perfectly
overlap each other. $M^*$ and $\alpha$ are correlated, and
the correlation is shown in Figure 6 of GMA99.

We check for a possible
incompleteness in the data sample, by computing the LF adopting
half a magnitude brighter limiting magnitude in $r$: the
newly determined best fit parameters are within the 68 \%
confidence level of the previous ones (i.e. the difference is
less than 1/$\sqrt{2}$ combined $\sigma$), in agreement
with independent checks done in GMA99.

The parameters just settled (i.e. $P=Q=0$) is the original Blanton
et al. (2001) choice for the first SDSS LF derivation, and also the 
standard setting adopted in most LF derivations, such as the 
Stromlo--APM LF (Loveday et al. 1992),
SSRS2 LF(da Costa et al. 1994), 
the CfA1 and CfA2 LF (Marzke et al. 1994), 
the Century Survey LF (Geller et al. 1997)
the Corona Borealis LF (Small et al. 1997)
the ESP LF (Zucca et al. 1997) and
the K20 LF (Pozzetti et al. 2003).

\subsection{$P=0$ and $\alpha$ fixed at the best global fit}

Let now $Q$ to be free (i.e. we allow $M^*$ to evolve), 
and $\alpha$  to be fixed (i.e. $P=0$)
at its best global fit. 
This setting is the same adopted in the Lin et al. (1999) LF determination.
Blanton et al. (2003) in their second SDSS LF derivation
do not use a Schechter function, but the sum of Gaussians, with fixed (i.e. not
evolving) relative amplitude. Therefore, also in Blanton et al. (2003)
the LF shape (slope) is not allowed to be redshift dependent.

Figure 2 shows confidence contours for $M^*$ and $\alpha$ for our sample. 
As expected, the present computed
$M^*$ differs from the one computed in the previous section, by 
$Q \times z_{median}$, as it occurs between the first and second SDSS LF derivations
(Blanton et al. 2003), the latter being different from the former ``almost
entirely because of the inclusion of evolution in the luminosity function model"
(Blanton et al. 2003). 

$Q$ turn out to be positive, i.e. galaxies are brighter in the past, in the three
filters. $Q=0$ is ruled out at $\sim$90 to 99.73 \% confidence
level in the $g$ filter only, and therefore the statistical significance
of the evidence is $\sim$2 to 3 $\sigma$ at most. 

For an old passive evolving stellar population, $Q=1.1-1.2$ at the redshifts
sampled in this work. Such values can easily derived from the convolution of the
GISSEL96 (Bruzual \& Charlot 1993) spectral energy distributions at the look back
times sampled in the present work (i.e. $z<0.25$) with the Gunn filters. Best fit
$Q$ values (2.1 and 3.1, see Table 1) derived by imposing a not evolving $\alpha$
are twice/three times larger than expected by
assuming a passive luminosity evolution, $Q=1.1-1.2$, but compatible with
them at the 68.3 (or better) \% confidence level (see Figure 2).

If we stop our analysis here we would claim to have found 
evidence of luminosity evolution at $\sim$2 to 3 $\sigma$, 
in good agreement with Blanton et al. (2003),
because our and their $Q$ values agree to better than $1 \sigma$. 
Instead, we proceed further.

\subsection{$Q$ forced to be as model predictions}

\begin{figure}
\psfig{figure=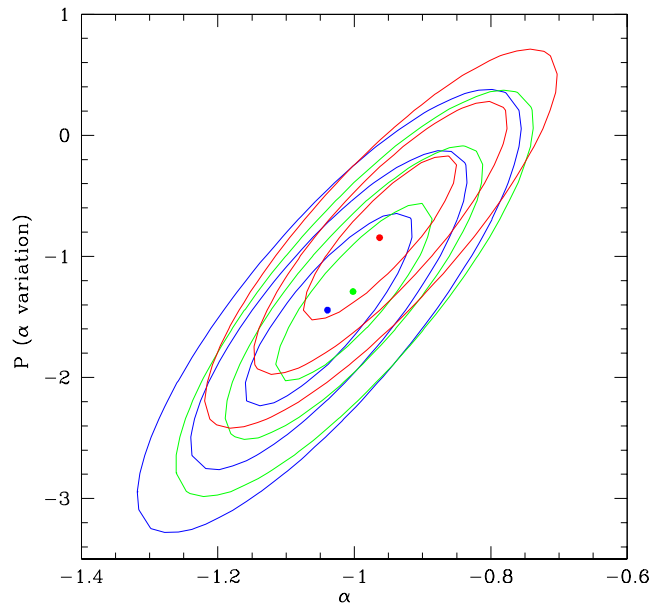,width=8.5truecm}
\caption[h]{Confidence contours at the 
68.3, 95.4 and 99.73 \% confidence levels 
for two degree of freedom, for $\alpha$ and $P=\partial \alpha / \partial z$. 
$M^*$ is forced to passively evolve. 
$g$, $r$ and $i$ contours are in blue, green and red
(also bottom to top).}
\end{figure}

We now assume that $Q$ is forced to be equal to the model prediction, i.e.
$Q=1.1-1.2$ depending on filter, and $\alpha$ free to evolve.  Figure 3 shows
confidence contours for $P$ and $\alpha$. The best fit $P$ (i.e. the $\alpha$
derivative) is $\sim -1$ to $-1.5$, although  $P \ne 0$ can be rejected at
only the $\sim 95$ \% confidence level (Figure 3). Therefore, $\alpha$
variations are suggested by the data but not required (evidence is 
$\sim 2 \sigma$), if passive evolution is
assumed. 

With the further constraint of a fixed $\alpha$, this type of fit 
would make resemblance to
the 2dFRS LF model (Norberg et al. 2002b), that assumes an evolution
well reproduced by simple synthesis population models, and a unique
(redshift independent) slope. Of course, being $M^*$ evolution
imposed ab initio, it cannot independently derived by the LF analysis.
At most, one may claim that the found luminosity evolution is 
compatible with the assumed one.

\begin{figure}
\psfig{figure=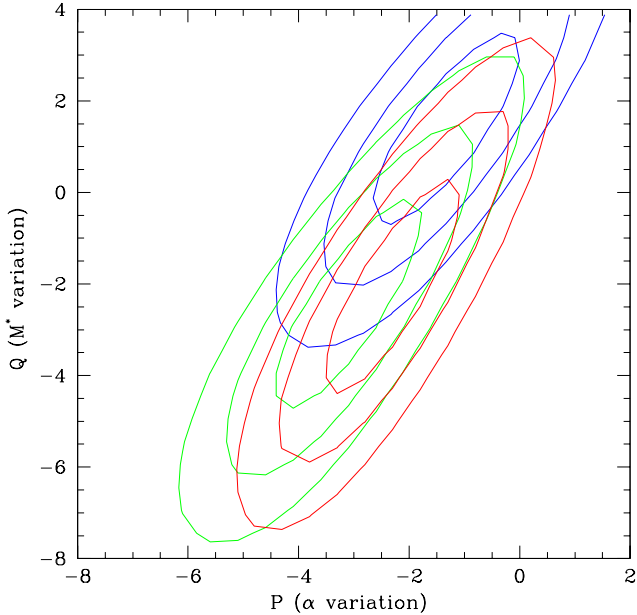,width=8.5truecm}
\caption[h]{Confidence contours at the 
68.3, 95.4 and 99.73 \% confidence levels 
for two degree of freedom, for $P= \partial \alpha
/ \partial z$ and $Q=\partial M^* / \partial z$ when
both $\alpha$ and $M^*$ are allowed to evolve. 
$r$, $i$ and $g$ contours are in green, red and blue
(also, left to right). The no--evolution locus is given
by $P=Q=0$. The passive evolving locus has $Q\sim1.1-1.2$.}
\end{figure}

\subsection{All parameters free}

Let now $P$ and $Q$ to be free, in order to look for evolution
in both the parameters. We do not force a priori that some of the parameters
do not evolve with redshift, but we left them to evolve as the data
allow. At difference of previous sections (and previous works,
including SDSS and 2dFRS ones), in absence
of a better knowledge we do not make a debatable assumption, and we allow
all our parameters to be redshift dependent, as much as
allowed by the data. We are here explicitly
allowing bright and faint galaxies to evolve differently, as it is
likely, because brighter galaxies tend to be redder and are therefore
presumed to be older and slowly evolving.

Figure 3 shows $P,Q$ confidence contours for the three filters.
The $P=Q=0$ point, i.e. the no--evolution case, is well outside the 99.73 \%
confidence contours in $g$ and  $r$, whereas falls on the top of the 99.73 \%
confidence contours in the $i$ band. Therefore, unevolving $\alpha$ and $M^*$ are
rejected at 3, or more, $\sigma$ in all three filters. Therefore, evolution is
required by the data, although it is not by adopting the other LF considered models. 

The point $Q\sim1.15$ and $P=0$, i.e. the pure passive evolution expectation, is
included in the 95.4 percent confidence contour in $g$ and $i$ and in the
99.73 percent confidence contour in $r$.
Therefore, the passive evolution case is  a statistically
acceptable description of the data at 2 to 3 $\sigma$.
Pure passive evolution, however, is not the unique solution
allowed by the data.  There are other acceptable solutions at  $Q\sim1.15$ (i.e.
$M^*$ is passively  evolving): a detailed inspection of Figure 3 shown  that all the
range $-3 \la  P \la 0$ is as good as (or better than) $P=0$. We remind
that negative values for $P$ imply that the LF becomes flatter at higher redshift. 

By leaving all parameter free, $M^*$ evolution is
smaller, or even with opposite sign, than by keeping $\alpha$ 
fixed (see Table 1). Therefore, {\it one more degree of freedom of evolution 
drastically influences the derived $M^*$ evolution}.
Such a degree of freedom is allowed by none of 
previous LF determinations, including 2dFRS and SDSS. Therefore,
we warn the reader that other similar studies in literature, all
of which assume a fixed and not evolving $\alpha$, may mis--interpret the
found $M^*$ evolution. 

By leaving all parameter free, errors on $M^*$ and $\alpha$ are larger (by
a factor $\sim3$) than when $P=Q=0$, i.e. under the usual hypothesis done in
the LF computation, and are twice larger than when $\alpha$ is kept fixed.

In order to check that the flattening is not due to an increasing incompleteness at
faint magnitudes with increasing redshift, we re--computed the best fit parameters
considering only the galaxies at least half a magnitude brighter than the claimed
completeness magnitude. We found almost identical best fit values (see Table 1
where we list results for the $r$ filter).  A couple of clusters, mainly nearby,
have deeper observation than the other clusters. In order to make the limiting
magnitude even more uniform with redshift, we recomputed the best fit parameters by
adopting the brighter between the claimed magnitude limits and $M_r=-17.5$ mag. We
found identical best fit values and errors (see Table 1), as expected because only
an handful of galaxies, out more than 2000, are removed by this cut. Therefore,
best fit values and confidence contours are robust to incompleteness.

\begin{table*}
\caption{Best fit parameters derived for different models}
\begin{tabular}{lccccl}
\hline
filter &  $M^*$ & $\alpha$ &  $Q \equiv \partial M^* / \partial z$
 & \hfill $P \equiv \partial \alpha / \partial z$ & Notes \\
\hline
\multispan{6}{no evolution ($Q=P=0$) \hfill} \\
 $r$  & $-21.38 \pm 0.08$ & $-0.84 \pm 0.04$ &  (0) &  (0) & \\
 $g$  & $-21.08 \pm 0.08$ & $-0.88 \pm 0.04$ &  (0) &  (0) & \\
 $i$  & $-21.72 \pm 0.07$ & $-0.85 \pm 0.04$ &  (0) &  (0) & \\
 $r$  & $-21.40 \pm 0.08$ & $-0.88 \pm 0.04$ &  (0) &  (0) & $mag_{lim}$ half mag brighter\\
\hline
\multispan{5}{$\alpha$ fixed, i.e. Evolution on $M^*$ only ($P=0$) \hfill} & see also Fig. 2 \\
 $r$  & $-21.02 \pm 0.11$ & (-0.82) & $2.1 \pm 0.7$ &  (0) & \\
 $g$  & $-20.53 \pm 0.11$ & (-0.83) & $3.1 \pm 0.8$ &  (0) & \\
 $i$  & $-21.46 \pm 0.12$ & (-0.84) & $1.5 \pm 0.8$ &  (0) & \\
\hline
\multispan{5}{$M^*$ passive evolving, and evolution on $\alpha$ only ($Q=0$) \hfill} & see also
Fig. 3 \\
$r$  & $-21.15 $ & $-1.00 \pm 0.08$ & (1.15) & $-1.3 \pm 0.6$  & \\
$g$  & $-20.84 $ & $-1.04 \pm 0.08$ & (1.20) & $-1.4 \pm 0.6$  & \\
$i$  & $-21.51 $ & $-0.96 \pm 0.08$ & (1.10) & $-0.8 \pm 0.6$  & \\
\hline

\multispan{6}{$M^*$ fixed, i.e. Evolution on $\alpha$ only ($Q=0$) \hfill} \\
 $r$  & $-21.01 $ & $-1.13 \pm 0.11$ & (0) & $-2.1 \pm 0.7$  & \\
 $g$  & $-20.53 $ & $-0.83 \pm 0.11$ & (0) & $3.1 \pm 0.8$  & \\
 $i$  & $-21.46 $ & $-0.84 \pm 0.12$ & (0) & $3.1 \pm 0.8$  & \\
\hline
\multispan{5}{$P$ and $Q$ free, i.e. evolution on both $M^*$ \& $\alpha$\hfill} & see also Fig. 4  \\
 $r$  & $-21.71 \pm 0.26$ & $-1.27 \pm 0.13$ & $-2.4 \pm 1.5$ & $-3.1 \pm 0.9$ &  \\
 $g$  & $-20.79 \pm 0.23$ & $-1.02 \pm 0.14$ & $ 1.4 \pm 1.4$ & $-1.3 \pm 0.9$ & \\
 $i$  & $-22.00 \pm 0.26$ & $-1.18 \pm 0.13$ & $-2.0 \pm 1.6$ & $-2.3 \pm 0.9$ & \\
 $r$  & $-21.67 \pm 0.28$ & $-1.24 \pm 0.15$ & $-2.1 \pm 1.7$ & $-2.7 \pm 1.6$ &
 $mag_{lim}$ half mag brighter \\
 $r$  & $-21.71 \pm 0.27$ & $-1.28 \pm 0.14$ & $-2.5 \pm 1.5$ & $-3.2 \pm 0.9$ & $max(mag_{lim},-17.5)$ \\
\hline
\end{tabular}

\hfill \break
All tabulated errors are 1 $\sigma$ one--parameter errors. See Figures
for two--parameters error contours. 
Fixed parameters are indicated in parenthesis.
\end{table*}

\subsection{$P=Q=0$ {\it vs} the correct solution when the LF evolves}

Using the very same data, GMA99 split the sample in two redshift bins and derived
the composed LF assuming no evolution inside each redshift bin ($\delta z \sim 0.1$). 
This is the standard way in which the evolution of the luminosity function is
computed (e.g. Lilly et al. 1995; Ellis et al. 1996; 
Heyl et al. 1997; Cohen et al. 2002). The
comparison of the LFs in the two redshift bins, performed using a $\chi^2$
approach, showed no statistically significant differences. In the present paper, by
adopting a model in which $\alpha$ do not evolve, 
we found an evidence of evolution on $M^*$ at 2
to 3 $\sigma$. A model that allow both $M^*$ and $\alpha$ to evolve excludes the
no--evolution case at better than 3 $\sigma$.

The analysis presented here supersedes the comparison presented in GMA99 
in two aspects: first, it 
uses the likelihood approach that is more powerful than the $\chi^2$.
Second, it removes
the logical inconsistency of checking for evolution LFs
derived {\it assuming} no evolution. 
The very same logical inconsistency is shared by
most LFs works aimed to measure an LF evolution (and finding an evolution), 
such
as CFRS (Lilly et al. 1995), Autofib
(Ellis et al. 1996, Heyl et al. 1997),
Caltech Faint Galaxy Redshift Survey (Cohen et al. 2002), 
and K20 (Pozzetti et al. 2003) LFs.

The assumption of an unevolving LF in each
redshift bin has an important impact on the luminosity density, even for
the modest (passive) evolution seen in the nearby universe:
the luminosity density derived in the $P=0$ hypothesis is 0.5 mag different from
the one derived if $M^*$ evolution is allowed (Blanton et al. 2003).
The impact of this
statement on the luminosity density evolution (the Madau plot) is 
obvious, in particular at high redshift,
especially when one realizes that evolution is faster at 
redshift higher than probed by SDSS, and that the 
redshift range covered by SDSS is smaller ($\delta z \la 0.2$)
than the typical redshift bin in usual luminosity density determinations
at high redshift.

\section{Conclusions and discussion}

There are been several comparisons of the cluster and field LFs in literature. 
However, almost none of them uses homogeneous and large samples as the ones used in this
paper, and none in the red bands considered in the present paper. 
Furthermore, only recently (Norberg et al. 2002b) there has been convergence
on the field LF.  Therefore, 
the results of previous cluster {\it vs} field LF
comparisons largely rely on which field LF has been considered. Here we
use the state-of-the-art LF, on which convergence seems to be reached.

In our cluster {\it vs} field comparison, the same LF model
is used in both environments: both are derived adopting $P=Q=0$,
and the two samples have very similar median redshifts, making the comparison
a fair one.  Cluster and field LFs, 
determined by analysing large homogeneous samples (SDSS, GMA99, PALal01), 
are almost indistinguishable down
to $M^*+4$, hence suggesting that
the effect of the cluster environment on the galaxy properties does
not affect the galaxy luminosity function in red filters. 
The similarity of the LF shape in different environments suggests that the galaxy
mass function is preserved during the galaxy infall in the cluster.

In order to measure the LF evolution
we use the sample presented in Garilli et al. (1996),
that is ideal from many points of view (multicolor data, large
size, many clusters, metric magnitudes). We use the same formalism (STY) 
used to compute many field LF, and we made different
assumptions on the model used to derive the LF, among which  
those adopted in recent LF determinations.

The found evolution depends on the LF model, i.e. whether the slope 
$\alpha$, or the characteristic magnitude $M^*$, or both, are 
allowed to evolve. Best fit parameters differ 
when different models are adopted,
in spite of the use of a fixed and large sample. 
Models in which both $M^*$ and $\alpha$ {\it may} evolve have not
been considered in previous studies and, therefore, the evidence
of a brightening of $M^*$ with look back time is intended by us as suggested
by previous works, but is still far to be definitively determined, 
even in the largest (and more recent) surveys.
Furthermore, errors depends on the model: errors are larger when more
degrees of freedom are allowed (or to be precise, when 
other parameters are not surreptitiously kept fixed because of a
lack of a better knowledge), hence suggesting that the usual errors
(derived adopting $P=Q=0$) are underestimated. The underestimation
is given by a factor $\sim1+n$, where $n=1,2$ are the number of parameters
kept fixed (see Table 1). This is one more reason to suggest caution.

In our more general
model, we found a statistically significant evidence of evolution, but the 
data do not clarify what ($M^*$, $\alpha$ or both) is actually changing.
Other works find a more constrained evolution, largely because
some evolutionary modes are not considered in these works.

One may argues that our results are due to the use of a poor sample,
or they concern only our sample, or they are derived using a peculiar method.  
First, our cluster sample is large, and
goes to $\sim M^*+4$ at $z\sim0.25$, whereas many of the field
surveys mentioned in this paper does not go that deep. Second, for the
LF determination we adopted the most common method used for field surveys.
Third, some or our results are confirmed on an independent
sample (and method): a better model (the addition of a monolithic 
evolution to an not evolving LF) 
changed the $M^*$ value by many $\sigma$ and the luminosity density by
half a magnitude (Blanton et al. 2003).
In our paper we show that an even better model, allowing a differential
luminosity evolution between bright and faint galaxies, changes the results 
once more. Fourth, some results (e.g. larger errors when adding LF degrees
of freedom) are expected to hold independently on the sample used.

Our discussion on the impact of the model choice on the found results has,
thus far, concerned almost exclusively LFs determined as described
in Sect. 3, i.e. by a maximum likelihood fit of the
Schechter function to the data (STY). Non parametric methods, 
such as step wise maximum likelihood (SWML), use a step function in
place of a Schechter one.  Since the problem we point out is a
logical one, and not a prerogative of the Schechter function, non 
parametric methods suffer in principle from similar problems
if the function is not explicitly allowed to evolve with redshift. 

The $1/V_{max}$ method (Schmidt 1968) is not immune to 
criticisms, too. Some field LFs, such as CFRS (Lilly et al. 1995), Autofib
(Ellis et al. 1996), K20 (Pozzetti et al. 2003), all were derived 
using the $1/V_{max}$ method (Schmidt 1968), 
i.e. weighing each galaxy by the reciprocal
of the maximal volume over which is observable. This method
assume a luminosity evolution in the $V_{max}$ computation. Therefore, 
when deriving the luminosity evolution from an LF computed
with the $1/V_{max}$ method, one
should claim, at most, that the found evolution is compatible with
the assumed one, without any guarantee that the solution is unique,
neither that the solution is the best one.

In the course of the analysis, we noted that there are 
logical inconsistencies in many field FL computations: 
in several papers dealing with the redshift evolution of the LF, the LF is
assumed not to evolve (inside each redshift bin in which it is computed)
during the LF computations, even when a luminosity evolution is 
found\footnote{As already mentioned, the model choice may badly affect best
fit values, errors and the luminosity density}.
Similarly, in several papers studying the environmental dependence
of the field LF, the LF is assumed to be environmental independent
during its computation. 

We suggest
that future studies solve at the same time for the LF parameters 
and their dependences (with look back time or environment).
Although we afforded the workload of re--computing our own cluster LF by
removing
logical inconsistencies, by allowing galaxies of different luminosity to
evolve independently each other, and by including an estimate of the
impact of the model assumptions on the final result, we left to 
other authors (disposing of all the needed data) to do a similar work 
for the numerous field LFs published thus far.

\medskip
When a Schechter function is adopted to describe the LF, $\alpha$
and $M^*$ (and their errors) are correlated, making the interpretation
of various findings less straightforward than for uncorrelated parameters.
Therefore, we suggest to break this correlation by adopting
a different $M^*$ definition not influenced by the LF slope
at faint magnitudes.
We propose to use a Petrosian--like definition of $M^*$, where the
characteristic magnitude $M^\eta$ is the magnitude at which the ratio between the
number of galaxies in the magnitude bin centered on $m$, $N(m)$ and in all
brighter magnitude bins, $N(<m)$, is equal to $\eta$ 

$$\eta = \frac{N(m)}{N(<m)}$$

For an opportune choice of $\eta$, $M^*\simeq M^\eta$ for a Schechter (1976) function.
Such a definition uses only the exponential cut--off of the LF, where $\alpha$ plays
a minor role (if $\eta$ is chosen to reproduce $M^*$).

We apply such an approach to our own data, for different choices of $\eta$ and 
of the bin width over which $N(m)$ is computed. Unfortunately, the application
of this approach do not clarify what is actually evolving
in our sample, mainly because it
uses only the bright part of the LF, which is also the less
populated in a cluster sample.
The proposed approach should be more efficient for a field
sample, 
because most of the field sample has magnitude brighter than (or similar to)
$M^*$, whereas for a cluster (volume limited) sample, most of
the galaxies are fainter than $M^*$. We propose that this approach is
tested on future LF determinations.

To summarize, we showed that the derived LF parameters depend on the
assumed model, in a way seldom recognized thus far. With our most
general model, that allows differential luminosity evolution between bright
and faint galaxies, 
we found a statistically significant evidence of evolution, but the data
do not unambiguously determine
what ($M^*$, $\alpha$ or both) is actually changing.  Adopting the
a unique model for large homogeneous samples, we found that cluster and field red 
LFs are quite similar suggesting a limited effect of the cluster 
environment on the galaxy
mass function.

\begin{acknowledgements}
The referee comments prompted us to better focus our work. We thank him.
\end{acknowledgements}

\appendix

\section{$Q=0$, i.e. un--evolving $M^*$}

For completeness, we also compute the best fit parameters in case
of a un--evolving $M^*$ (i.e. $Q=0$). Literature papers do not
explore the possibility of an evolving $\alpha$, for lack
of data.

Inspection of the $\alpha-P$
confidence contours (figure not shown) in the three filters
shows that $P=0$ can be rejected at
the 95.4 \% confidence level at most. Therefore,
for this model assumptions, 
we have no compelling evidences for an $\alpha$ variation.

\end{document}